%% file: 0_Main.tex
\documentclass[conference]{IEEEtran}
\IEEEoverridecommandlockouts
\usepackage{cite}
\usepackage{algorithm}
\usepackage{amsmath,amssymb,amsfonts}
\usepackage{algorithmic}
\usepackage{graphicx}
\usepackage{textcomp}
\usepackage{xcolor}
\usepackage{verbatim}
\usepackage{diagbox}
\usepackage{soul}
\usepackage{subfigure}
\def\BibTeX{{\rm B\kern-.05em{\sc i\kern-.025em b}\kern-.08em
    T\kern-.1667em\lower.7ex\hbox{E}\kern-.125emX}}
\renewcommand{\baselinestretch}{0.965}    
\begin{document}

\title{A Novel Dynamic Bandwidth Allocation Design for 100G Coherent Passive Optical Network}
\author{\IEEEauthorblockN{ Rujia Zou\IEEEauthorrefmark{1}, Haipeng Zhang\IEEEauthorrefmark{2}, Karthik Sundaresan\IEEEauthorrefmark{2}, Zhensheng Jia\IEEEauthorrefmark{2}, Suresh Subramaniam\IEEEauthorrefmark{1}}
\textit{\IEEEauthorrefmark{1}The George Washington University},
\big\{rjzou, suresh\big\}@gwu.edu\\
\textit{\IEEEauthorrefmark{2} Cable Television Laboratories, Inc},
\big\{h.zhang, k.sundaresan, s.jia\big\}@cablelabs.com\\
}

\maketitle

\begin{abstract}
With the rapid advancements in coherent Passive Optical Network (PON) technologies featuring 100G and higher data rates, this paper addresses the urgent requirement for sophisticated simulation and MAC layer development within the domain of coherent Time Division Multiplexing (TDM) PON and coherent Time and Frequency Division Multiplexing (TFDM) PON networks. The ever-growing demand for latency-sensitive services and expanding user populations in next-generation 100G and beyond coherent PONs, underscores the crucial need for low-latency bandwidth management and efficient Dynamic Bandwidth Allocation (DBA) mechanisms. In this paper, we present a pioneering analysis of two established DBAs from the perspective of temporal misalignments. Subsequently, a novel DBA algorithm tailored for coherent PONs featuring 100 Gbps data rate and up to 512 end-users is introduced, named the Hybrid-Switch DBA. This innovative approach allows for adaptive switching of the DBA scheme in response to real-time traffic conditions. To the best of our knowledge, this paper represents the first attempt to address the misalignment problem of DBA and proposes a novel DBA solution for both TDM- and TFDM-based coherent PON networks. This research significantly contributes to the development of coherent TDM PON and coherent TFDM PON networks by enhancing the efficiency of bandwidth allocation and addressing the challenges associated with misalignments in DBA mechanisms. As optical access networks continue to evolve to meet the ever-increasing demands of modern communication services, the Hybrid-Switch DBA algorithm presented in this paper offers a promising solution for optimizing network performance and accommodating latency-sensitive applications.

\end{abstract}


\input{1_Introduction}

\input{2_Motivation}

\input{3_System}

\input{4_DBA}

\input{5_Simulation}

\input{6_Conclusion}

\input{DBA.bbl}



\end{document}

%% file: 1_Introduction.tex
\section{Introduction}
The pursuit of delivering reliable bandwidth with minimal network latency remains a primary focus within optical access networks. The ever-expanding landscape of emerging services, including 5G mobile X-haul, edge computing, AR/VR gaming, Tactile Internet, and UHD video distribution, has ushered in additional demands upon access networks \cite{1,2}. These demands emphasize the growing importance of low latency and high reliability in future access networks, as they are expected to cater to increasingly latency-sensitive services. In response, the development of new deterministic and reliable latency management approaches becomes imperative. For instance, stringent latency requirements range from 1 to 10 milliseconds (ms) for the F1 mobile fronthaul interface, dwindling to mere hundreds of microseconds (within 1ms) when transitioning to a lower-layer function-split of mobile fronthaul \cite{1}.

The Passive Optical Network (PON), a point-to-multi-point system, has historically stood as a dominant architecture for enabling bandwidth sharing across diverse service types \cite{3, 4}. Dynamic Bandwidth Allocation (DBA) plays a pivotal role in PONs by efficiently allocating upstream bandwidth based on real-time user demands from optical network units (ONUs) and network congestion conditions. The effectiveness of DBA algorithms or strategies profoundly influences both upstream bandwidth efficiency and latency performance \cite{zhu2020pwc,5}. Nevertheless, the majority of existing DBA schemes are tailored for relatively small user populations. Traditional PON DBA designs are optimized to manage bandwidth sharing for 32 users per PON port in EPON or GPON systems \cite{skubic2009comparison}. These designs typically entail the OLT polling all users to gather bandwidth requests before resource allocation. However, when user counts exceed 64, such as 128, 256, or even 512, the existing DBA approach introduces substantial latency overhead, particularly in Time Division Multiple Access (TDMA) environments.

\begin{figure}[ht]
\center
\includegraphics[scale=0.22]{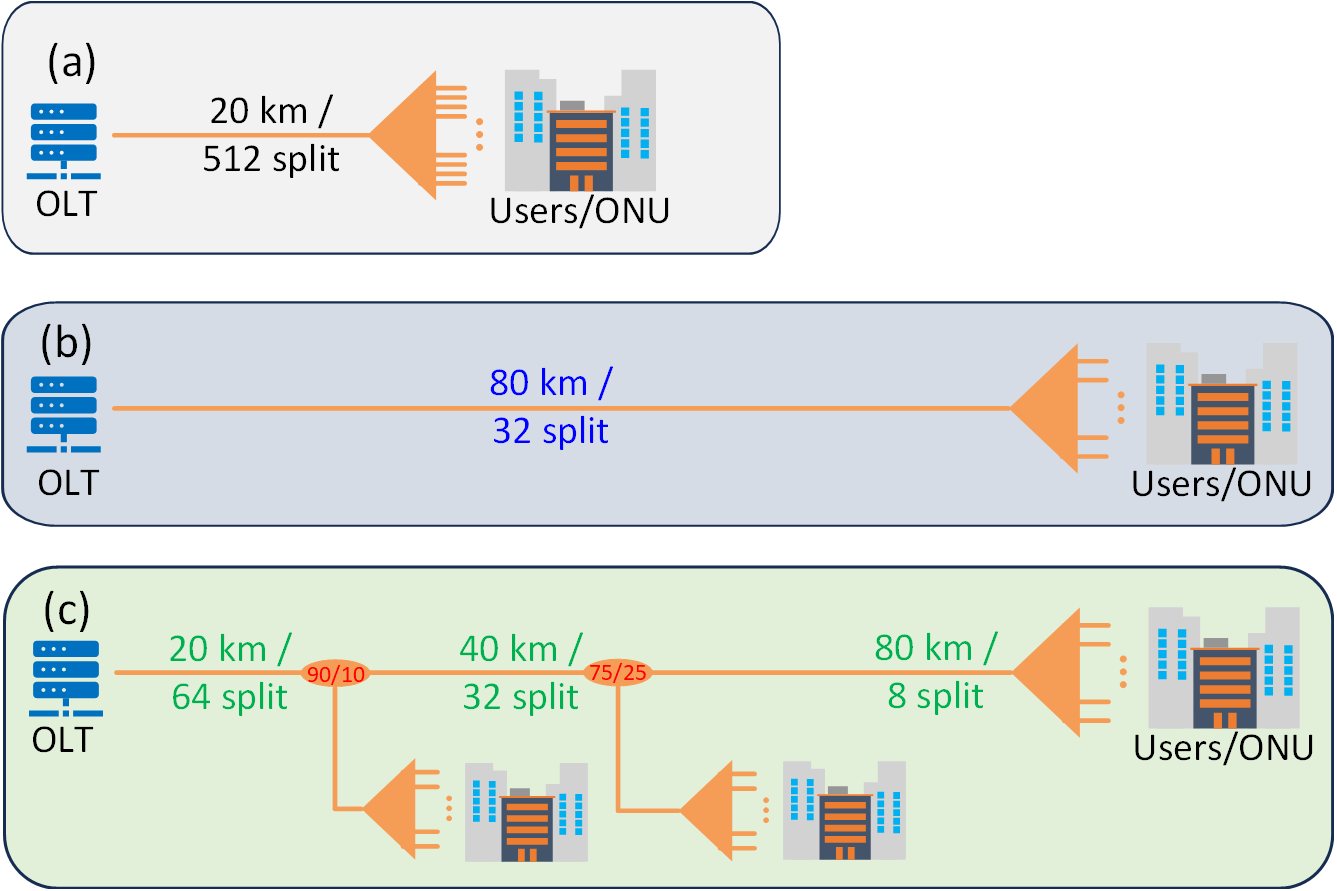}
\caption{Coherent PON deployment architectures.}
\label{Architectures}
\end{figure}

Addressing this latency challenge and optimizing PONs to support larger user populations has spurred research into reducing guard interval (GI) overhead. Coherent PON (CPON) emerges as a promising solution, boasting superior receiver sensitivity and ample capacity to accommodate up to 512 users on a single wavelength using Time Division Multiplexing (TDM) \cite{zhang2021efficient,6, 7}. This calls for a versatile DBA architecture and control process capable of accommodating large split ratios (e.g., 1:128, 1:256, and 1:512) and providing flexibility to serve user populations ranging from small to large (from 32 to 512), as depicted in Fig. \ref{Architectures}. With its substantial power budget, CPON excels in supporting a significant number of user connections with short fiber transmission distances, particularly in densely populated areas, e.g., 512 users supported by a 20km fiber link, as illustrated in Figure 1(a). Moreover, it possesses the capability to extend services to users located at a distance from the Hub, even with small split ratios, as exemplified in Figure 1(b), where 32 users are supported with an 80km fiber transmission \cite{8}. Furthermore, a hybrid deployment scenario, as shown in Figure 1(c), allows for dynamic combinations of distance and split ratios. Therefore, the need arises for a DBA scheme with multiple modes, offering flexibility across various deployment scenarios and the ability to accommodate varying service-level agreement requirements to support forthcoming low-latency services with adaptable quality of service (QoS).

Another compelling solution in the realm of coherent PON is TFDM, leveraging digital subcarrier (DSC) multiplexing technology \cite{8,9}. TFDM integrates TDM and frequency division multiplexing (FDM) by allocating different frequency bands to each optical signal \cite{7}. Each frequency band carries an independent data stream, and time slots within each band serve to transmit data from different users or services. TFDM not only offers substantial bandwidth optimization but also provides flexibility in latency management and capacity allocation \cite{xu2022intelligent}.

TFDM's inherent coherence and selection filtering capabilities enable direct application within power splitter-based Optical Distribution Networks (ODN). TFDM-based coherent systems partition data into multiple narrower intermediate DSCs, which are subsequently modulated onto the optical carrier in a parallel fashion. TFDM eliminates the need for colored optics and allows for flexible allocation of frequency bands to network services or groups of ONUs with distinct latency or capacity requirements. This innovative approach significantly reduces scheduling latency and traffic blocking rates, facilitating efficient bandwidth utilization, flexible link budget designs, and scalable architecture \cite{8}. Coherent TFDM technology, with its enhanced flexibility and capabilities, emerges as a compelling candidate for future access networks.

Considering these developments, this paper underscores the pressing need for advanced simulation and MAC layer development in the context of coherent PON and coherent TFDM PON networks. The pursuit of low-latency bandwidth management and efficient DBA mechanisms is paramount to meeting the evolving demands of optical access networks in an era of latency-sensitive services and growing user populations. In this paper, we first present a novel analysis of two classic DBAs from a 'misalignments' perspective. Then we propose a new DBA algorithm for coherent PON, named Hybrid-Switch DBA, where the DBA is switched according to the traffic conditions. To the best of our knowledge, this is the first paper that considers the misalignment problem of DBA and proposes a new DBA accordingly for both TDM- and TFDM-based coherent PON. 

The rest of the paper is organized as follows. The motivation is presented and analyzed in Section~\ref{sec: Motivation}. The system model and problem statement are defined in Section~\ref{sec: system}. The new DBA algorithm is presented in Section~\ref{sec:DBA}. Simulation results are given in Section~\ref{sec: results}, and the paper is concluded in Section~\ref{sec:conc}.

%% file: 2_Motivation.tex
\section{Motivation}
\label{sec: Motivation}
In this section, we demonstrate the motivations of this work. 
\subsection{DBA for Coherent PONs}
As the landscape of passive optical networks continues to evolve, the development of coherent PONs represents a significant forward. Coherent PONs have emerged as a solution to meet the escalating demands of modern communication networks, offering higher data rates and accommodating an ever-expanding number of ONUs. This remarkable enhancement in network capacity and throughput, however, brings forth novel challenges in the context of efficient resource allocation and QoS optimization. Traditional DBA approaches, well-suited for earlier PON, GPON, and XGPON configurations, are now faced with the imperative need for adaptation to the unique attributes of coherent PONs. The necessity for a novel DBA framework tailored to this advanced PON architecture is evident, as it aims to harness the full potential of these networks while addressing the surging number of ONUs and the heightened network capacity. 

\subsection{Misalignments in DBA}
In the realm of PONs, DBA plays a pivotal role in efficiently managing and distributing available bandwidth among ONUs. However, a critical issue that has thus far received limited attention is the noticeable time gap between the OLT receiving queue size information from ONUs and the actual execution of DBA within the ONUs. This temporal disparity gives rise to a distinctive challenge in PONs, one that pertains to the synchronization between network monitoring and network management. This temporal misalignment, which can significantly impact the network's performance and QoS, remains an understudied aspect of PONs. Recognizing this gap in the existing body of research, this paper sets out to address this issue by proposing a novel algorithm designed to rectify this temporal incongruity and thereby enhance the overall efficiency of DBA in PON.

\subsection{TFDM}
Furthermore, as we delve into the DBA challenges within TFDM-based PONs, it becomes evident that the adoption of the TFDM techniques holds great promise but has remained largely unexplored in the existing research of PON and DBA. TFDM is an innovative multiplexing technique that offers a unique approach to optimizing the allocation of resources within PONs. Unlike traditional methods, TFDM facilitates the transmission of data over both time and frequency dimensions, offering enhanced flexibility and resource utilization. One intriguing aspect of TFDM-based DBA that merits attention is its potential to address the diverse QoS requirements of different groups of ONUs. By leveraging TFDM, PONs can assign specific DSCs to different groups of ONUs, thereby tailoring the bandwidth allocation to their specific QoS needs. This capability allows for the provisioning of dedicated channels for high-priority or 'VIP' groups of ONUs, ensuring that their stringent latency requirements are met. This allocation strategy not only enhances overall network performance but also maximizes customer satisfaction, making TFDM-based DBA a compelling avenue for further investigation.

%% file: 3_System.tex
\section{System model}
\label{sec: system}
In this section, we introduce the system model of this work, including the network mode and the DBA cycle.

We consider the PON model the same as a conventional PON, where a set of ONUs are given and connected to the OLT by a splitter. ONUs are located at the customer premises and serve as endpoints for data transmission. They receive downstream data and send upstream data. The OLT serves as the central point of control in the PON, managing the network's downstream and upstream traffic. With TFDM, the upstream and downstream channels are divided into several DSCs with predefined partitions. Each DSC is assigned to its predetermined ONU group.  

\begin{figure}[ht]
\center
\includegraphics[scale=0.30]{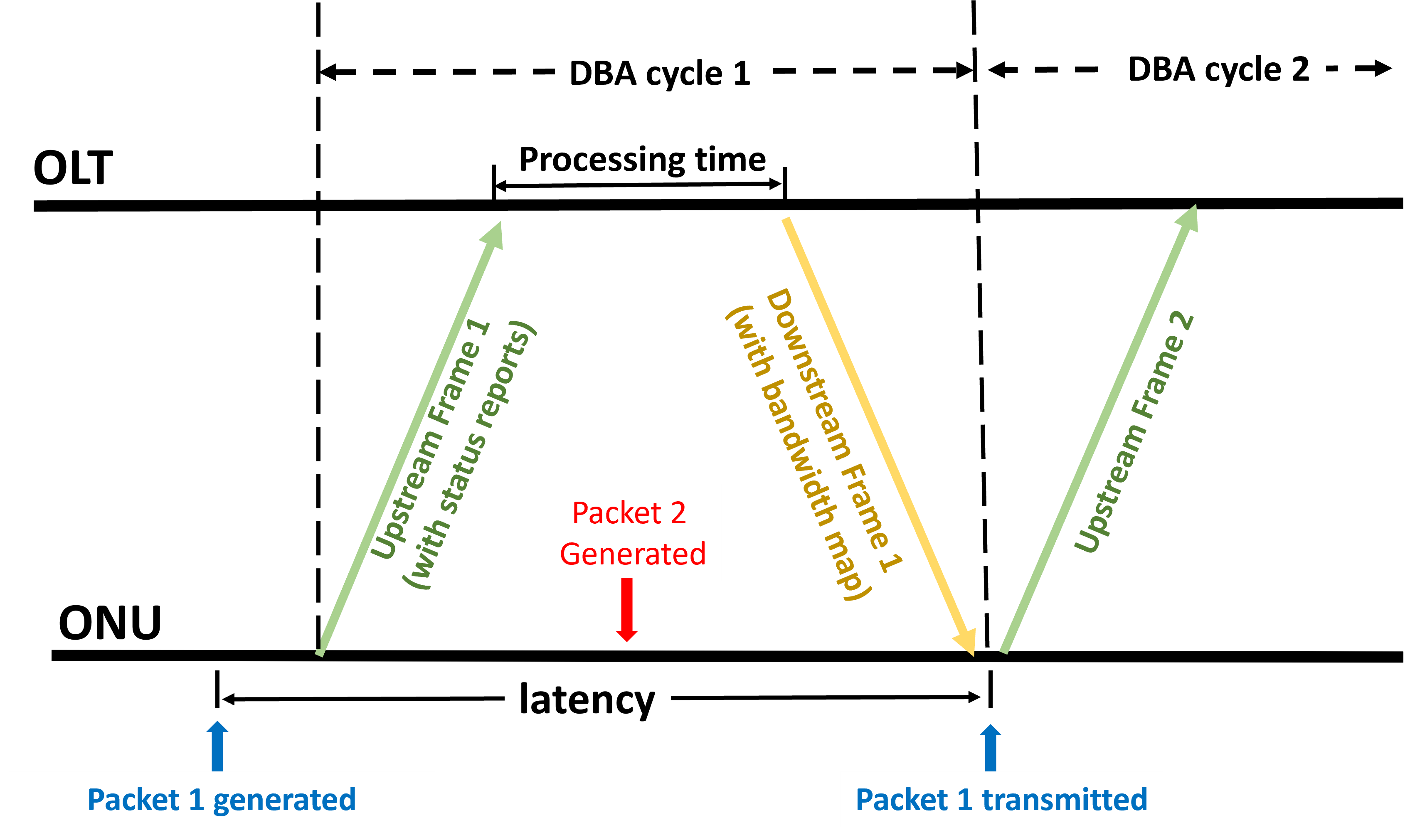}
\caption{DBA cycle.}
\label{DBA cycle}
\end{figure}

An example of the DBA cycle is shown in Fig. \ref{DBA cycle}. The DBA cycle begins with the continuous monitoring of data queues at each ONU. Each ONU tracks the size of its transmit queue, representing the data that is awaiting transmission to the OLT. In a status-reporting DBA, ONUs periodically report their queue sizes and network conditions to the OLT in the upstream. These status reports provide the OLT with the queue information to make informed decisions about resource allocation. Once the OLT receives the upstream with the status reports, the OLT starts to process the DBA and calculates the bandwidth map (BWmap). During the process, the traffic keeps generating traffic packets at the same time (e.g. packet 2 in the example). The BWmap carries the transmission start time and grant size for each ONU for the next cycle. The BWmap is broadcasted to all the ONUs along with the downstream. Once the ONUs receive the downstream frame, each ONU transmits its data during its allocated time slots based on the granted size and starts a new DBA cycle. .

We assume that the packets are generated as the user data for each ONU. Those packets are stored in the queue of each ONU before the transmission. The latency of a packet is defined as the time gap between packet transmission time and packet generation time. The objective of the DBA is to minimize the average latency of all the packets. 

%% file: 4_DBA.tex
\section{Dynamic Bandwidth Allocation}
\label{sec:DBA}

\subsection{Round-Robin}
The Round-Robin (RR) DBA is a commonly used algorithm in PONs for the allocation of upstream bandwidth resources to ONUs. This algorithm is designed to distribute available bandwidth uniformly among ONUs, ensuring that all ONUs receive equal treatment and opportunities to transmit data, promoting fairness in the network.

However, in high network load scenarios, RR may lead to inefficient resource allocation, causing network congestion and increased average latency. Inappropriate decisions during congestion can accumulate very fast due to the high network load, degrading the general QoS and leading to an inevitably congested network condition. Meanwhile, when ONUs have low data or even no data to transmit during their allocated time slots, bandwidth resources can be wasted. This can lead to an inefficient utilization of available bandwidth.

\subsection{Weighted-Fair}
The Weighted-Fair (WF) DBA is an adaptive algorithm used in PONs to allocate upstream bandwidth resources based on the queue size of each ONU. ONUs with larger queues are granted more bandwidth to transmit their data, while ONUs with smaller queues receive proportionally smaller grants. WF seeks to strike a balance between fairness and efficiency to lower the average latency. It ensures that ONUs with higher data loads have a greater opportunity to transmit their data, preventing potential bottlenecks and avoiding potential large latency caused by the in-balance traffic pattern. It also avoids over-allocating resources to ONUs with lower data loads, thereby improving overall network efficiency.

However, with the consideration of the DBA process time, WF does not address bursty traffic effectively. If some ONUs occasionally have short but high-priority bursts of data, WF may not provide them with the appropriate resources in a timely manner due to the misalignment we mentioned in section \ref{sec: Motivation}. The WF decision is based on the queue size but the queue size changes very fast along with the bursty traffic. Suppose a large number of packets are generated from the ONUs between two upstream frames, since the BWmap is calculated based on the queue size information carried by the previous frame but executed after the generation of those packets, the new incoming packets have to wait for the next cycle to be transmitted. When the traffic load is low, the misalignment is more severe since the bursty traffic can dramatically change the network traffic.  

\subsection{DBA analysis}
\begin{figure*}[htbp]
    \centering
    \subfigure[]
    {
        \includegraphics[scale=0.22]{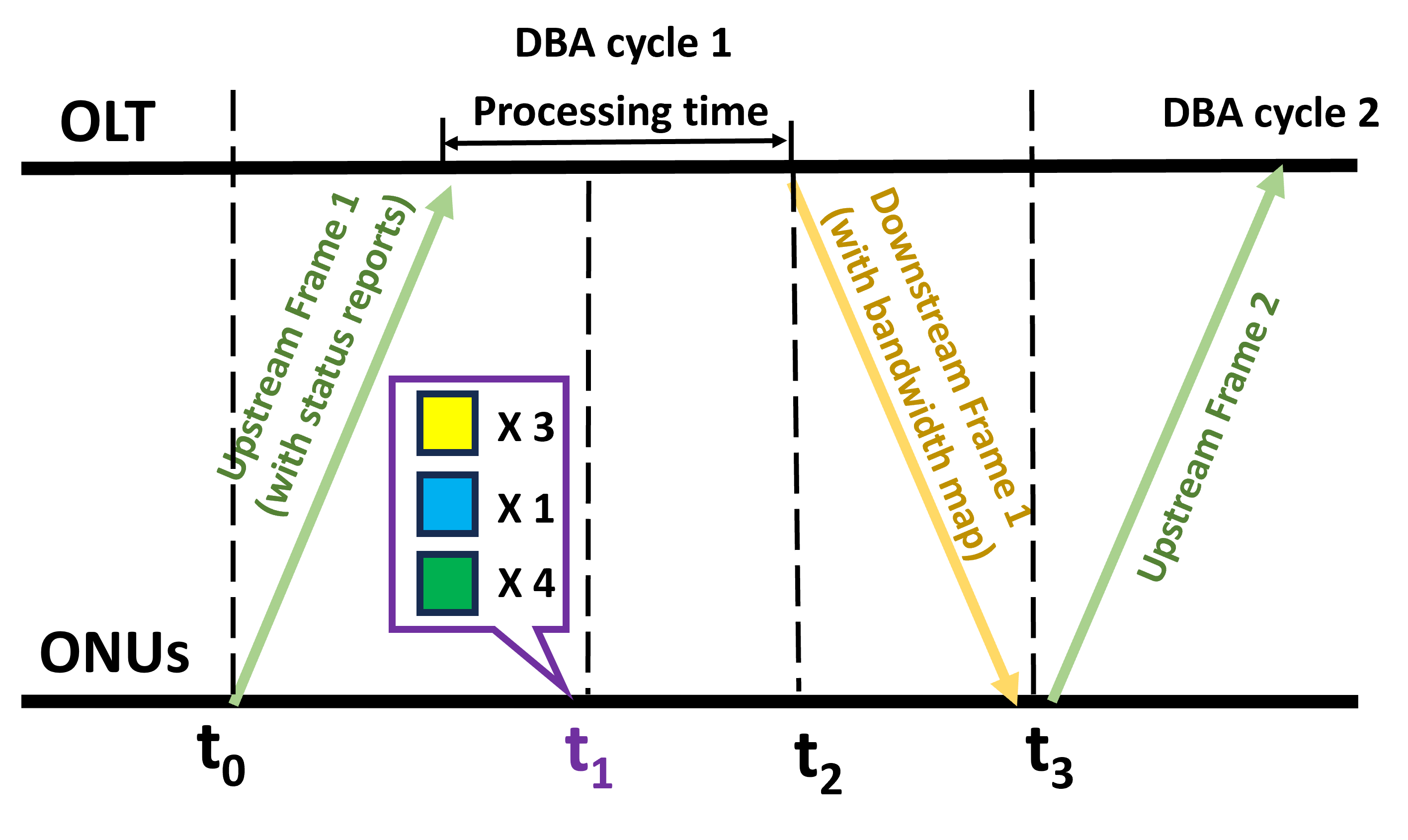}
        \label{DBA_E2}
    }
    \subfigure[]
    {
        \includegraphics[scale=0.22]{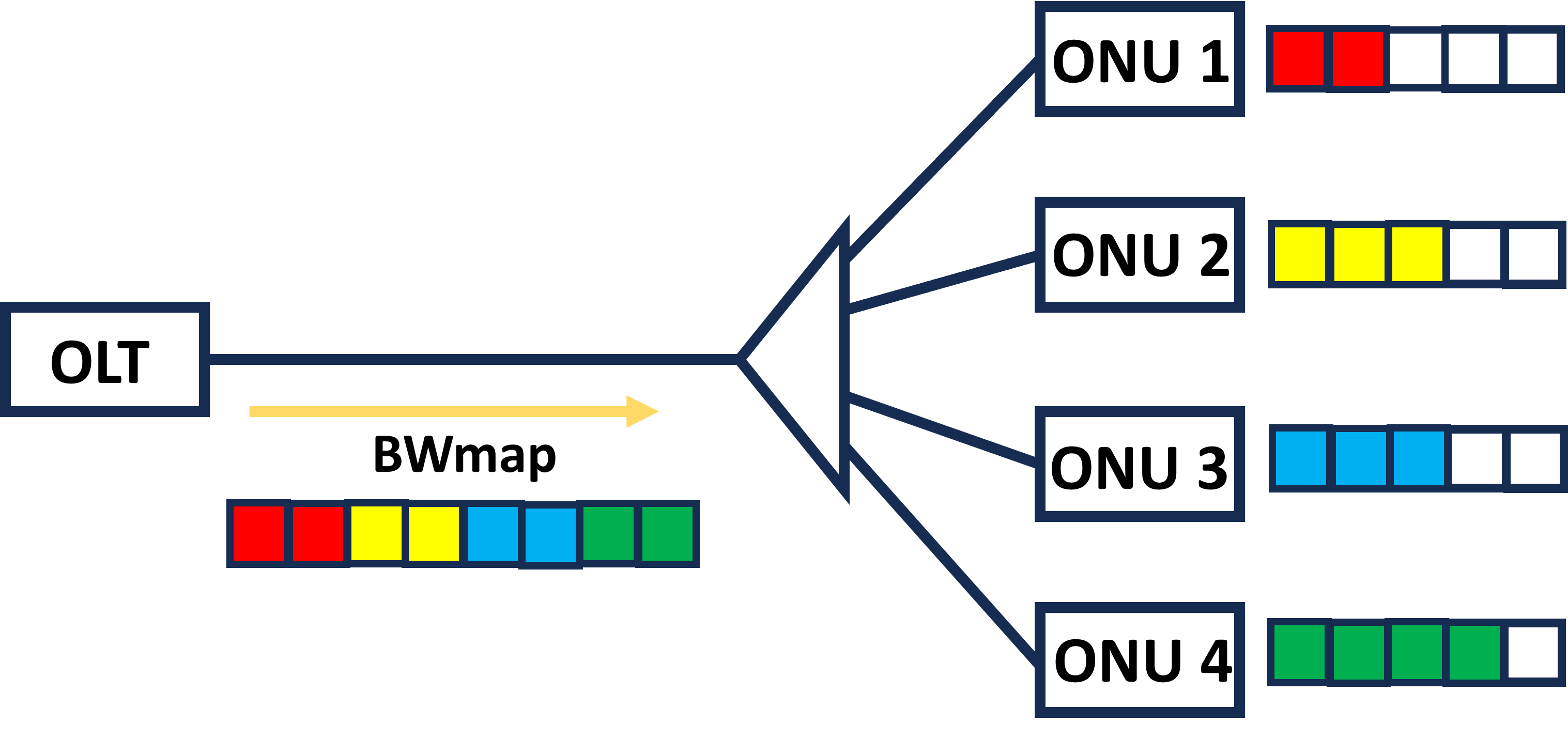}
        \label{DBA_RR1}
    }
    \subfigure[]
    {
        \includegraphics[scale=0.22]{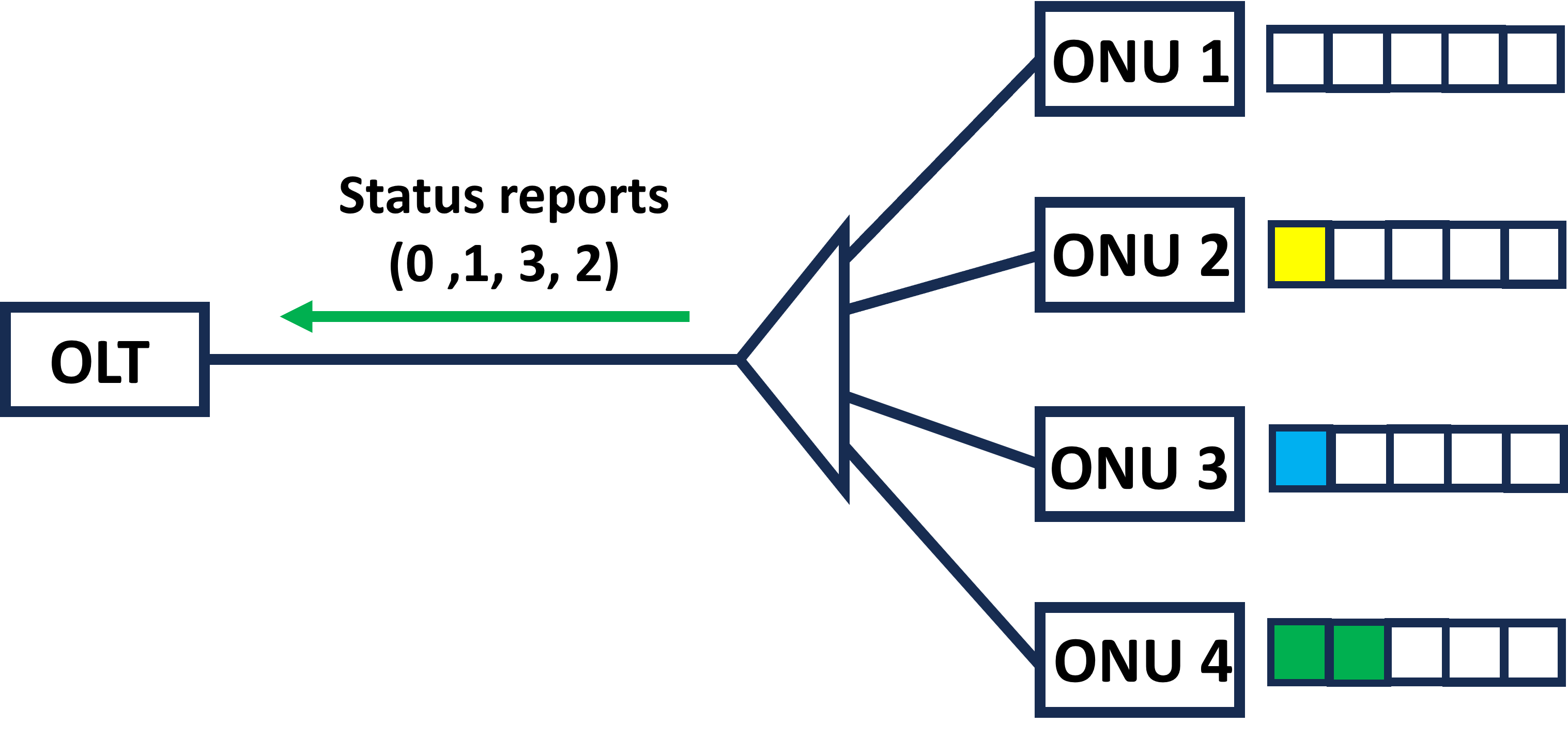}
        \label{DBA_RR2}
    }
    
    \subfigure[]
    {
        \includegraphics[scale=0.22]{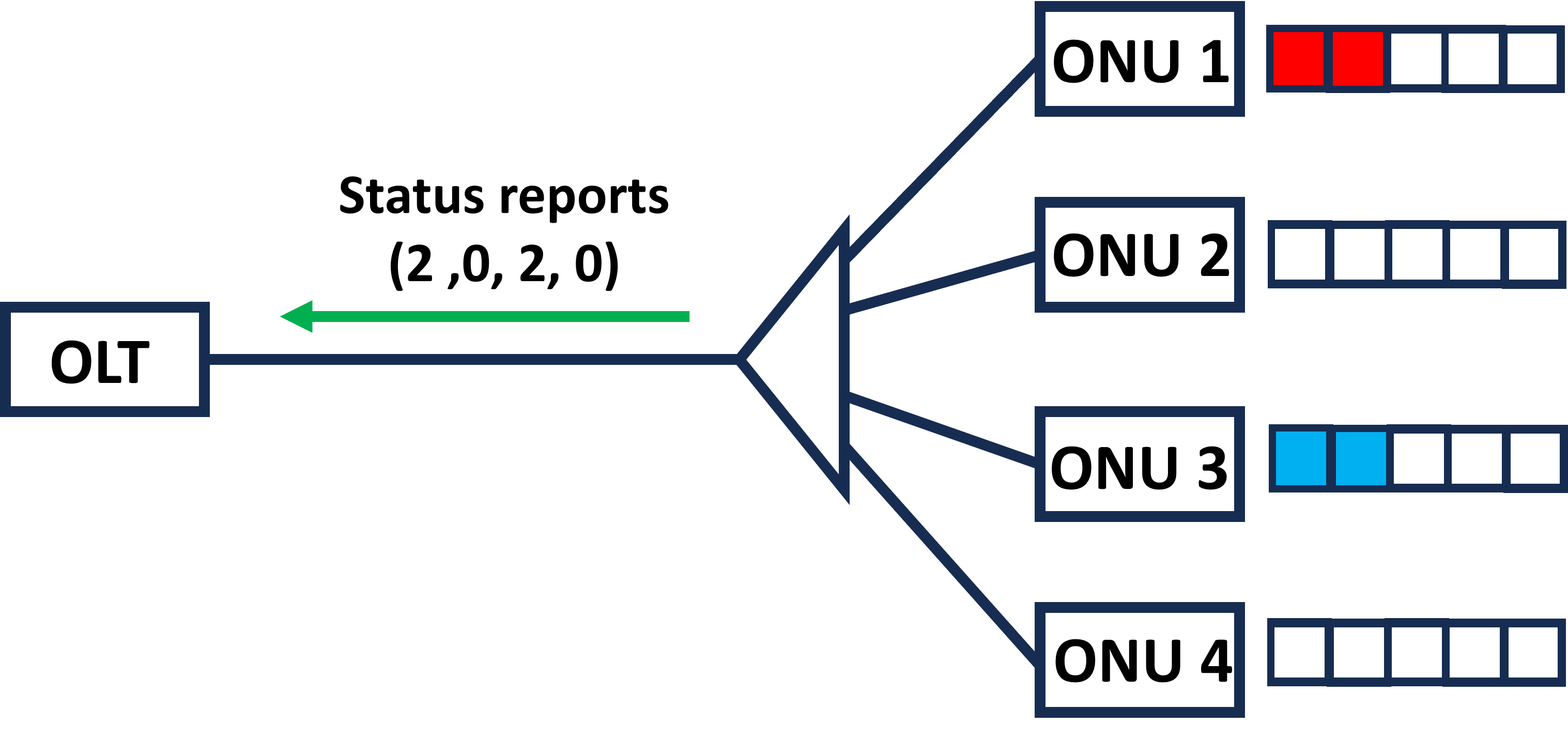}
        \label{DBA_E1}
    }
    \subfigure[]
    {
        \includegraphics[scale=0.22]{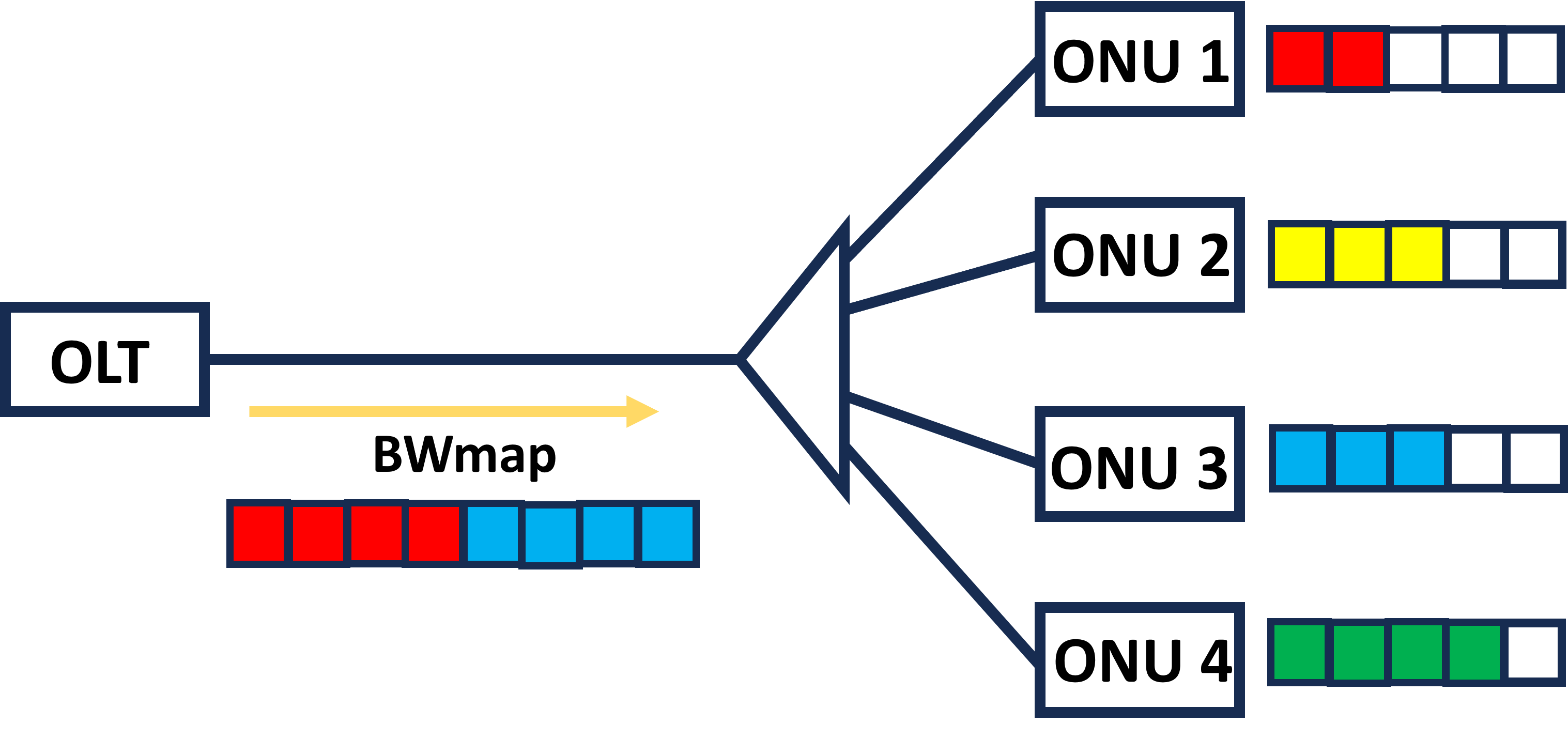}
        \label{DBA_WF1}
    }
    \subfigure[]
    {
        \includegraphics[scale=0.22]{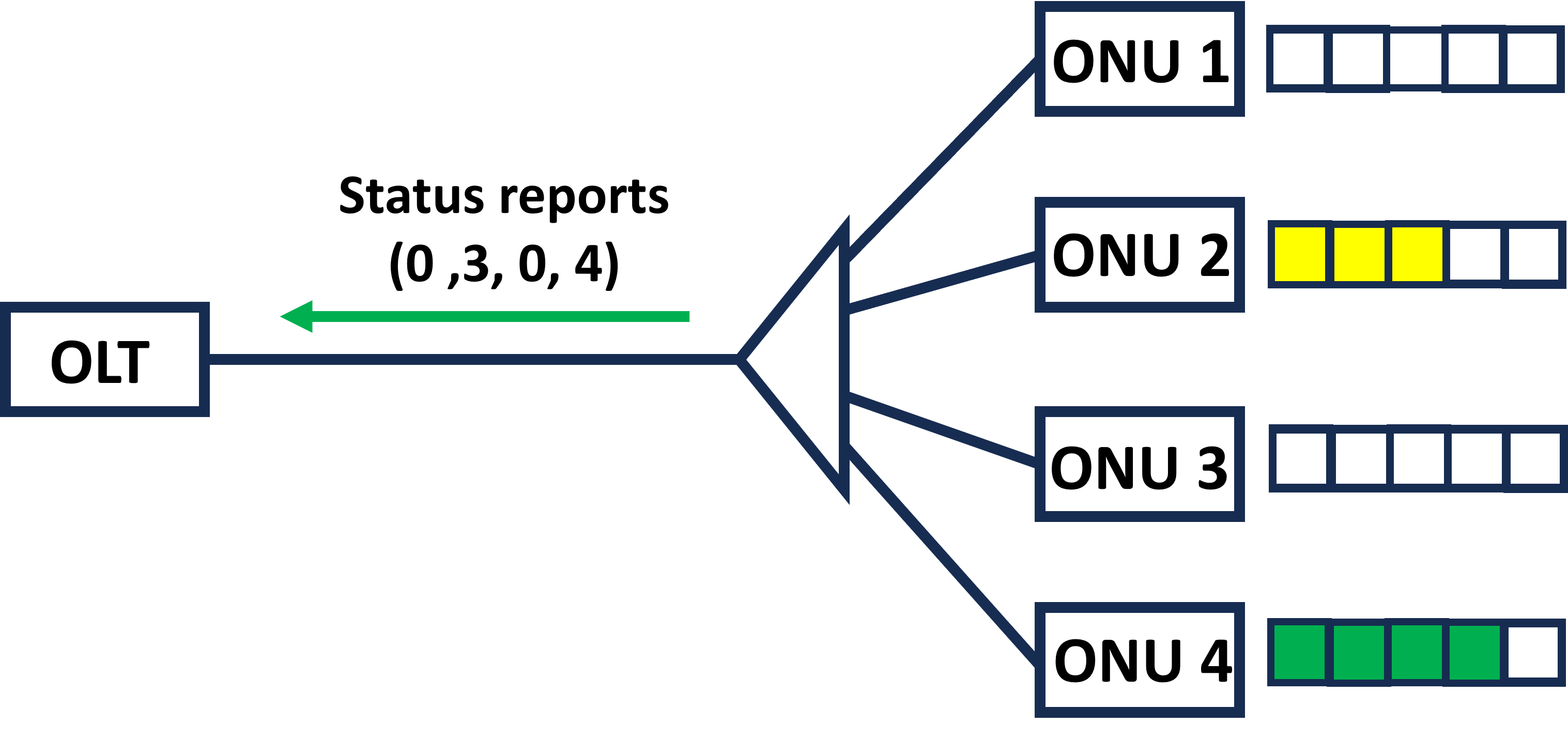}
        \label{DBA_WF2}
    }
    
    \caption{Examples of RR and WF dynamic bandwidth allocation.}
    \label{DBA_example}
\end{figure*}

In order to further demonstrate the features of the two DBAs above, and analyze the misalignment of the DBA, examples for RR and WF are shown in Fig. \ref{DBA_example}.  

The time diagram of the examples is shown in Fig. \ref{DBA_E2} and the initial state of the network (when $t = t_0$) is shown in Fig. \ref{DBA_E1}. We can see that there are two packets waiting in ONU 1 (marked in red) and two packets waiting in ONU 3 (marked in blue). ONU 2 and ONU 4 are idle. Suppose we use the number of waiting packets to measure the queue size of each ONU, the status reports are (2, 0, 2, 0), which are the numbers of packets in ONUs. The status reports are sent to the OLT for DBA calculation. The BWmaps calculated by OLT are shown in Fig. \ref{DBA_RR1} and Fig. \ref{DBA_WF1} for RR and WF respectively. Here we assume that each upstream frame is able to carry 8 packets. We can see that in the BWmap of RR, the bandwidth is equally assigned to 4 ONUs, even though the queue size values of ONU 2 and ONU 4 in status reports are both 0. In the BWmap of WF, the bandwidth is shared by ONU 1 and ONU 3 according to the reports. During the DBA process at time $t_1$, we assume that there are several packets just generated by ONUs, which are shown in the box in Fig. \ref{DBA_E2}. There are 3 packets from ONU 2, 1 packet from ONU 3 and 4 packets from ONU 4. The new queue conditions for ONUs are updated and shown in Fig. \ref{DBA_RR1} and Fig. \ref{DBA_WF1}. In \ref{DBA_RR2} and Fig. \ref{DBA_WF2}, the upstream frame is formed according to two BWmaps, where $t = t_3$.  

From these examples, we can see that even though the RR DBA does not consider the queue size of the ONU, the new packets are transmitted on time. On the other hand, the WF makes decisions based on the reports. However, due to the misalignment of the DBA, more packets waiting for the next round. The reason is that the new packets generated during the DBA process time are considered by the OLT. The reports provided by the ONUs at time $t_0$ are not 'Fresh' when its corresponding BWmap is executed at time $t_3$. We can see from these examples that WF is not always better than RR as we expected. RR could be a better option when the misalignment is bad. Also, we cannot overlook the fairness and efficiency of WF DBA. 

Meanwhile, with the TFDM technology, PONs can assign specific channels to different groups of ONUs. The traffic patterns and subcarrier capacities may vary among different groups. For the different subcarrier capacity to traffic load ratios, the DBA should be changed to cover the misalignment issue. Therefore, a new DBA design is necessary to balance the advantages and drawbacks of these two classic DBAs. 

\subsection{Hybrid-Switch DBA}
\begin{figure}[ht]
\center
\includegraphics[scale=0.5]{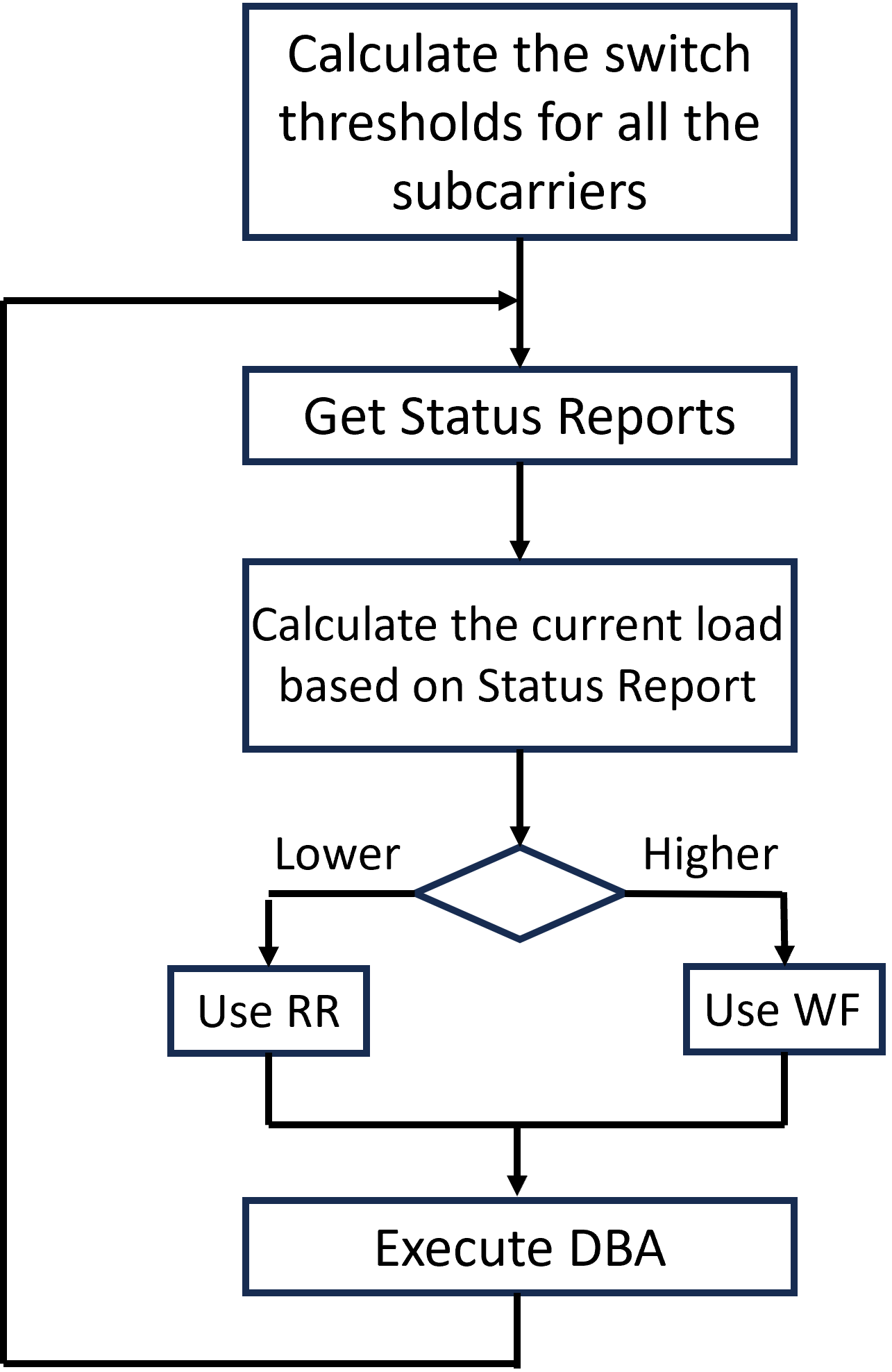}
\caption{The procedures of the Hybrid-Switch DBA.}
\label{new_dba}
\end{figure}

Based on the fact we mentioned, now we present our new DBA algorithm. In this innovative DBA approach, the network employs a hybrid mechanism that adapts to varying network load conditions by seamlessly transitioning between RR DBA and WF DBA. 
The procedures are shown in Fig. \ref{new_dba}.

The choice of DBA mode—whether to use RR or WF is determined by a threshold function designed to assess the current network load. We assume that the network load is measured in terms of the total number of bytes of the waiting packets. If it is below the defined threshold for low load, RR is employed. In this mode, ONUs are granted bandwidth on a round-robin basis, promoting simplicity and predictability. On the other hand, when the network load is higher than the defined threshold for high load, the system switches to WF. WF allocates bandwidth based on the queue sizes of ONUs, allowing for efficient resource utilization and fairness in scenarios with higher network demand. This adaptive DBA approach ensures that the network remains responsive to real-time needs, maintains fair allocation, and optimizes resource utilization, regardless of the prevailing load conditions.

In this work, we assume that the threshold is calculated. The threshold is defined as follows:
\begin{equation}
threshold = \alpha \times \frac{C}{N}
\end{equation}

$\alpha$ is the threshold factor, $C$ is the transmission capacity of each DBA cycle (125 microseconds) of the subcarrier or channel, and $N$ is the number of ONUs assigned to the subcarrier. The current load is calculated according to the status reports and compared to the threshold. In our work, the $\alpha$ is set to 1.5. Suppose the bandwidth of a channel is 25 Gbps and the transmission capacity is 25 Gbps $\times$ 125 microseconds. The idea of this design is that if the total amount of bytes per ONU is close to or even lower than the amount of bytes that can be transmitted in one upstream frame, then the misalignment will highly affect the performance of DBA as we discussed in Section \ref{sec:DBA}.C.

%% file: 5_Simulation.tex
\section{Simulation Setup and Results}
\label{sec: results}
\subsection{Simulation Setup}
Now we present the performance evaluation and analyses of the proposed Hybrid-Switch DBA scheme. In order to fully highlight the improvement of the Hybrid-Switch DBA design over the two classic DBAs in terms of the packet blocking ratio, we implemented a software network simulator for upstream DBA in coherent PONs. 

In the PON system, we assumed that there is 1 OLT and N ONUs. For each ONU, the round trip time (RTT) is randomly selected between 80 to 120 microseconds. An unlimited buffer is deployed on each ONU to store the waiting packets. All the waiting packets are sorted in terms of the arrival time. Every ONU belongs to a pre-determined user group.

For the traffic, packets are dynamically generated from ONUs according to a Poisson process with different arrival rates. The size of the packet ranges from 64 to 1518 bytes. In order to simulate the traffic pattern closer to real traffic from network users, we assume that the traffic generation rates of ONUs are changed during the simulation. The 'busy hour' is created jointly/dis-jointly for ONUs. The busy hours are created every $[P_{min}, P_{max}]$, and last $[L_{min}$, $L_{max}]$. During the busy hour, the packet arrival rate increases with a busy hour ratio $b$. In this work, we set the $P_{min}$, $P_{max}$, $L_{min}$, and $L_{max}$ to 2000, 3000, 500, and 1000 microseconds. 

The upstream channel is set to test two scenarios, one is symmetrical 100-Gb/s/$\lambda$ single carrier TDM coherent PON, while the other is symmetrical 100-Gb/s/$\lambda$ TFDM coherent PON based on four subcarrier multiplexing schemes. The physical layer synchronization block for upstream (PSBu) size, XG-PON encapsulation method (XGEM) header size, and the upstream burst guard time are set to 24, 8, and 8 bytes respectively \cite{konstadinidis2018multilayer}. 

\subsection{Results}
\begin{figure}[ht]
\center
\includegraphics[scale=0.5]{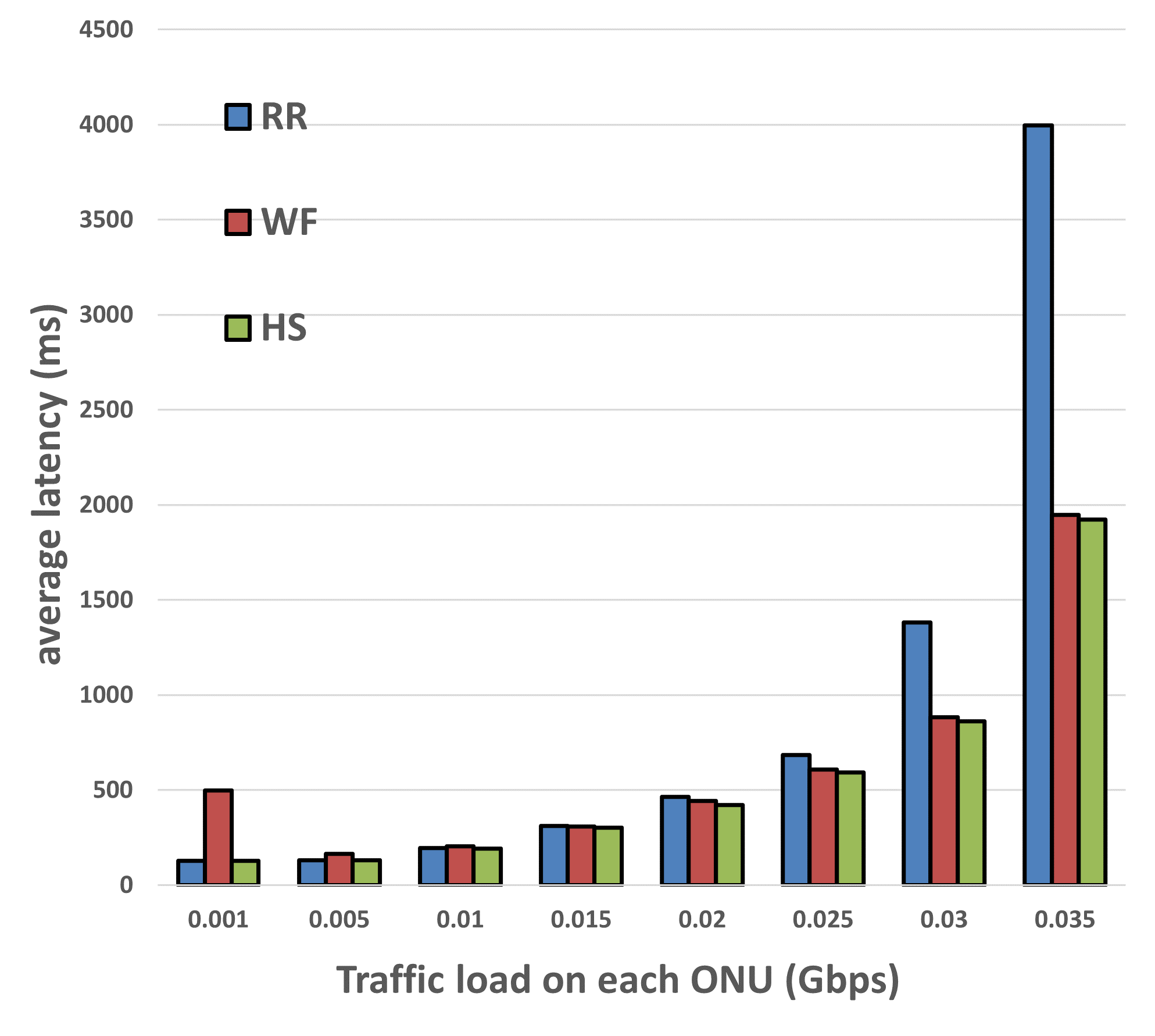}
\caption{The average latency of DBAs in single carrier scenario with different network loads.}
\label{results_1}
\end{figure}

First, we show the results for a single carrier TDM coherent PON case, which has a 100 Gbps capacity. The number of ONUs is set to 512 to simulate the large number of users in coherent PON. All the ONUs are set into a single group. Different ONUs have different busy hours. The average latency is shown in Fig. \ref{results_1}.

We can see from the results that, the latency of RR DBA is lower than the WF when the network load on each ONU is low. The reason is that the traffic variation can dramatically change the general traffic pattern and lead to the misalignment of DBA. When the network load is very large, the WF is better than the RR, because the fairness of the DBA dominates the performance. WF can come up with a better allocation where all the ONUs are assigned with bandwidth on demand.

Meanwhile, we can see that our proposed algorithm HS is able to switch between RR and WF accordingly and take advantage. When the network load is relatively low, RR is selected by HS, while the WF is activated by HS when the network load is higher.

\begin{figure}[ht]
\center
\includegraphics[scale=0.5]{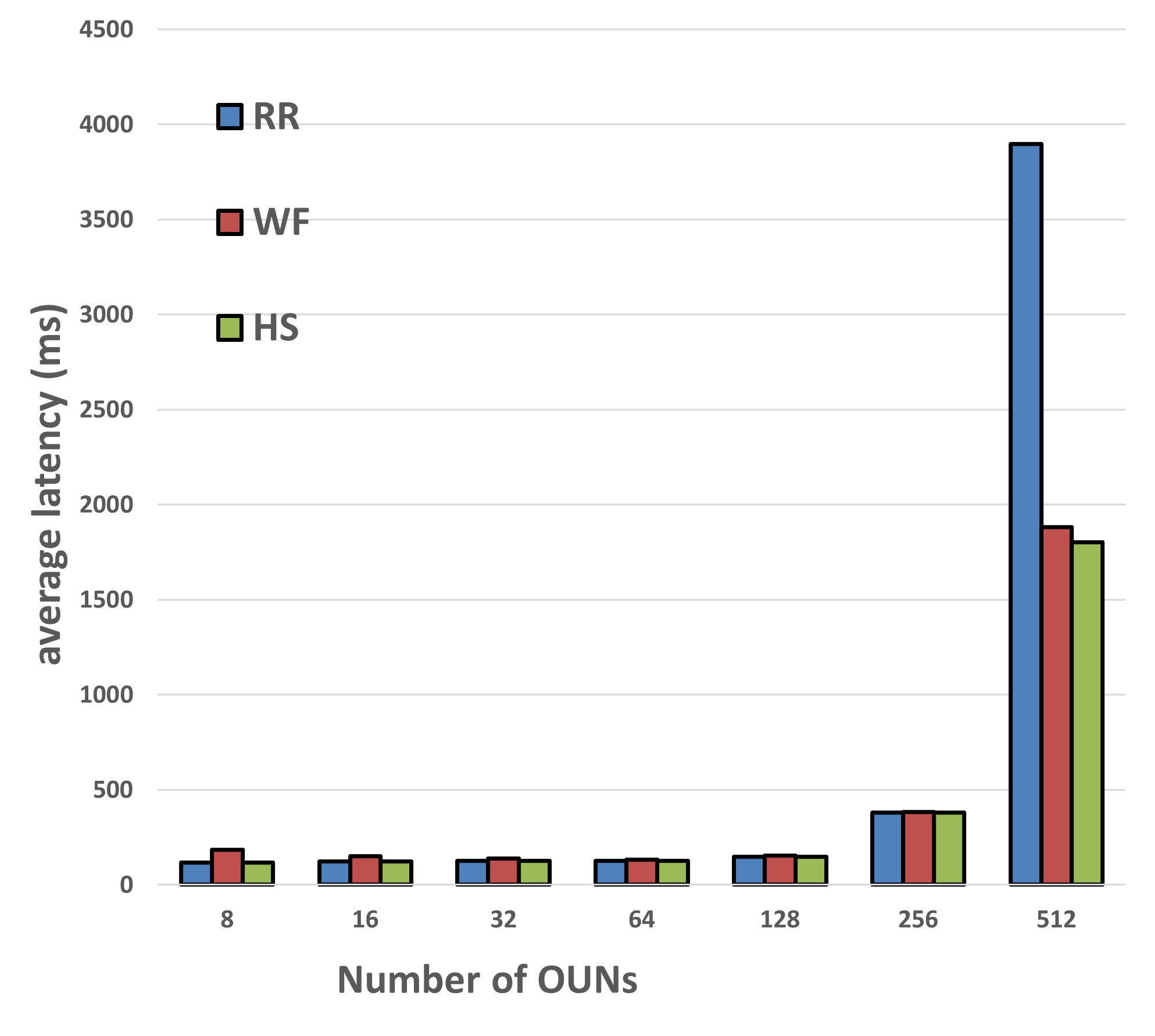}
\caption{The average latency of DBAs in single carrier scenario with different numbers of ONUs }
\label{results_2}
\end{figure}

In Fig. \ref{results_2}, we show the average latency of DBAs with different numbers of ONUs. Here we set the basic traffic load of each ONU without busy hour as 0.035 Gbps. First We can see that the average latency of RR is lower than WF when the network load is low, while WF performs much better than RR when the network load is high. Then, since the network loads are low when the numbers of ONUs are limited, WF has a larger average latency due to the misalignment. Same as the results we have in Fig. \ref{results_1}, WF is much better than RR when the network load is high. In addition, HS is always able to select the appropriate DBA to be used. 

Then we show the performance of RR, WF, and HS when they are deployed with TFDM consideration. We assume that there are four subcarriers, Ch1, Ch2, Ch3, and Ch4, each carrying 25Gbps. The ONUs are divided into two groups with the different class levels. The 'first class' group has 32 ONUs while the 'normal class' group has 480 ONUs. In order to guarantee the service of the first-class group, we assume that subcarrier Ch1 is reserved for the first-class group only. Ch2 to Ch4 is reserved for the normal-class group. In this case, the bandwidth reserved for each ONU in the first-class group is larger than the normal-class group. Therefore, the traffic loads of different groups are not the same, which leads to different scales of the misalignment problem. We assume that the basic traffic load for each ONU is 0.03 Gbps. The busy hour settings are the same as the previous TDM cases. 
\vspace{-4pt}
\begin{figure}[htbp]
    \centering
    \subfigure[]
    {
        \includegraphics[scale=0.6]{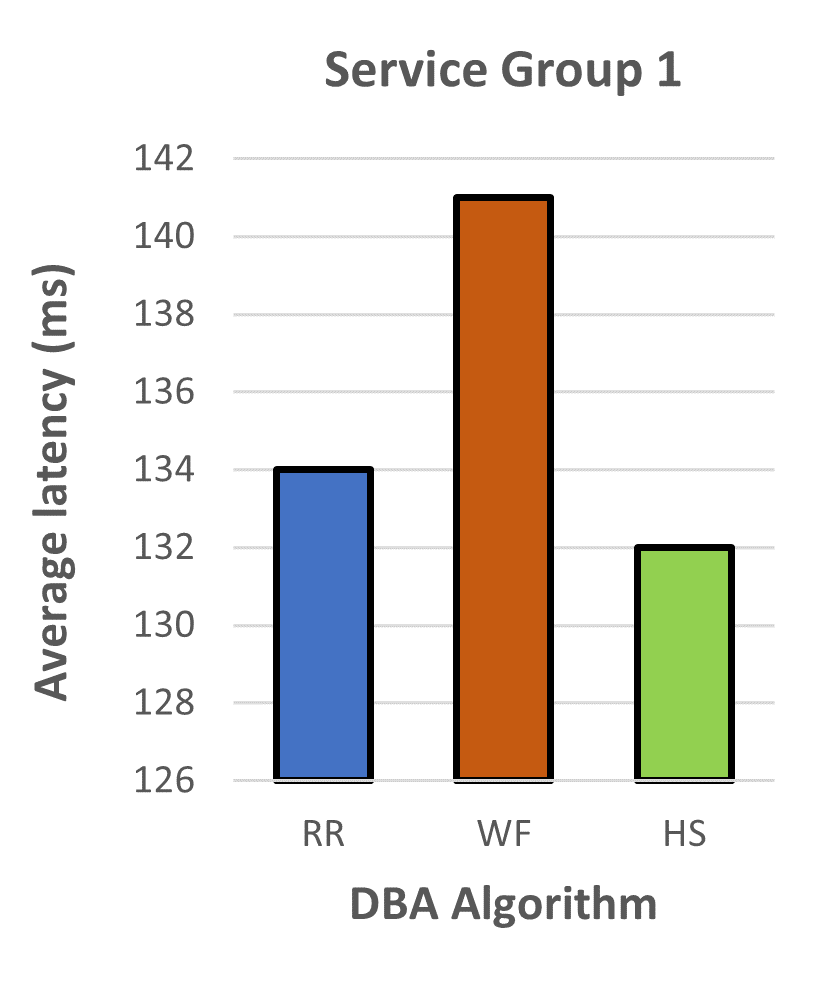}
        \label{results_3}
    }
    \subfigure[]
    {
        \includegraphics[scale=0.6]{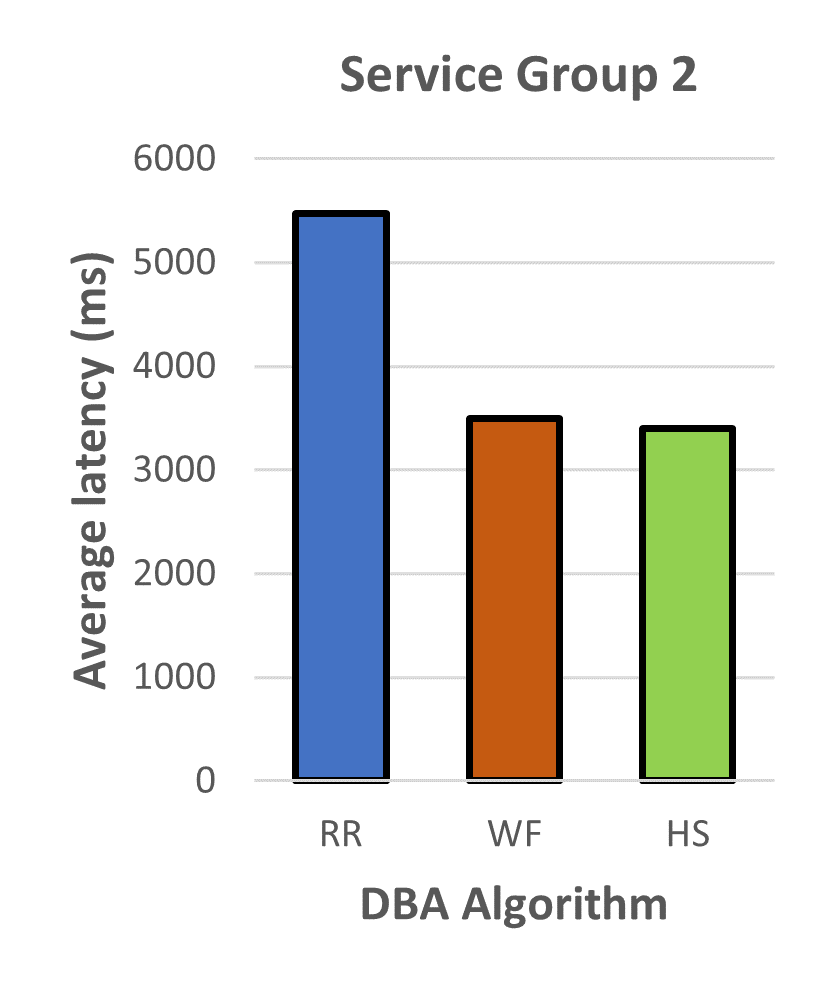}
        \label{results_4}
    }
    \caption{The average latency of two groups in the two-subcarrier scenario.}
    \label{subcarrier}
\end{figure}
\vspace{-4pt}

The average latency values in microseconds of the two-subcarrier cases are shown in Fig. \ref{subcarrier}. We can see that the latency of group 1 is very low and the HS DBA is working very well since its latency is very close and even lower than the RR. Group 1 has a very large bandwidth but limited traffic load so the misalignment happens badly. For group 2, we can see that HS is able to activate the WF since the traffic load is very high. The average latency results here show that the HS is able to recognize different traffic patterns for different groups and overcome the misalignment problem. 

%% file: 6_Conclusion.tex
\section{Conclusion}
\label{sec:conc}
In conclusion, this paper has highlighted the critical importance of advancing simulation and MAC layer development in the rapidly evolving field of coherent TDM PON and coherent TFDM PONs. The growing demand for latency-sensitive services and the continuous expansion of user populations in the next-generation coherent PONs underscore the crucial need for efficient DBA mechanisms and low-latency bandwidth management. This study has introduced a novel perspective on the analysis of two well-established DBAs, focusing on temporal misalignments. Subsequently, we have proposed the Hybrid-Switch DBA algorithm, which enables adaptive switching of the DBA scheme in response to real-time traffic conditions and offers an innovative solution to optimize network performance and accommodate the specific requirements of latency-sensitive applications. This study forms a robust basis for advancing and enhancing coherent PON technologies, securing their prominent position in the next-generation efficient, adaptive optical access networks.

%% file: DBA.bbl

%% file: 0_Main.bbl
\begin{thebibliography}{10}
\providecommand{\url}[1]{#1}
\csname url@samestyle\endcsname
\providecommand{\newblock}{\relax}
\providecommand{\bibinfo}[2]{#2}
\providecommand{\BIBentrySTDinterwordspacing}{\spaceskip=0pt\relax}
\providecommand{\BIBentryALTinterwordstretchfactor}{4}
\providecommand{\BIBentryALTinterwordspacing}{\spaceskip=\fontdimen2\font plus
\BIBentryALTinterwordstretchfactor\fontdimen3\font minus \fontdimen4\font\relax}
\providecommand{\BIBforeignlanguage}[2]{{%
\expandafter\ifx\csname l@#1\endcsname\relax
\typeout{** WARNING: IEEEtran.bst: No hyphenation pattern has been}%
\typeout{** loaded for the language `#1'. Using the pattern for}%
\typeout{** the default language instead.}%
\else
\language=\csname l@#1\endcsname
\fi
#2}}
\providecommand{\BIBdecl}{\relax}
\BIBdecl

\bibitem{1}
T.~Pfeiffer, ``Considerations on transport latency in passive optical networks,'' in \emph{45th European Conference on Optical Communication (ECOC 2019)}.\hskip 1em plus 0.5em minus 0.4em\relax IET, 2019, pp. 1--3.

\bibitem{2}
D.~van Veen and V.~Houtsma, ``Strategies for economical next-generation 50g and 100g passive optical networks,'' \emph{Journal of Optical Communications and Networking}, vol.~12, no.~1, pp. A95--A103, 2020.

\bibitem{3}
Z.~Jia and L.~A. Campos, \emph{Coherent Optics for Access Networks}.\hskip 1em plus 0.5em minus 0.4em\relax CRC Press, 2019.

\bibitem{4}
F.~Effenberger, D.~Cleary, O.~Haran, G.~Kramer, R.~D. Li, M.~Oron, and T.~Pfeiffer, ``An introduction to pon technologies [topics in optical communications],'' \emph{IEEE Communications Magazine}, vol.~45, no.~3, pp. S17--S25, 2007.

\bibitem{zhu2020pwc}
M.~Zhu, J.~Gu, and G.~Li, ``Pwc-pon: an energy-efficient low-latency dba scheme for time division multiplexed passive optical networks,'' \emph{IEEE Access}, vol.~8, pp. 206\,848--206\,865, 2020.

\bibitem{5}
H.~Uzawa, K.~Honda, H.~Nakamura, Y.~Hirano, K.~Nakura, S.~Kozaki, A.~Okamura, and J.~Terada, ``First demonstration of bandwidth-allocation scheme for network-slicing-based tdm-pon toward 5g and iot era,'' in \emph{Optical Fiber Communication Conference}.\hskip 1em plus 0.5em minus 0.4em\relax Optica Publishing Group, 2019, pp. W3J--2.

\bibitem{skubic2009comparison}
B.~Skubic, J.~Chen, J.~Ahmed, L.~Wosinska, and B.~Mukherjee, ``A comparison of dynamic bandwidth allocation for epon, gpon, and next-generation tdm pon,'' \emph{IEEE Communications Magazine}, vol.~47, no.~3, pp. S40--S48, 2009.

\bibitem{zhang2021efficient}
J.~Zhang, Z.~Jia, M.~Xu, H.~Zhang, and L.~A. Campos, ``Efficient preamble design and digital signal processing in upstream burst-mode detection of 100g tdm coherent-pon,'' \emph{Journal of Optical Communications and Networking}, vol.~13, no.~2, pp. A135--A143, 2021.

\bibitem{6}
R.~Matsumoto, K.~Matsuda, and N.~Suzuki, ``Fast, low-complexity widely-linear compensation for iq imbalance in burst-mode 100-gb/s/$\lambda$ coherent tdm-pon,'' in \emph{2018 Optical Fiber Communications Conference and Exposition (OFC)}.\hskip 1em plus 0.5em minus 0.4em\relax IEEE, 2018, pp. 1--3.

\bibitem{7}
J.~Zhang, Z.~Jia, M.~Xu, H.~Zhang, L.~A. Campos, and C.~Knittle, ``High-performance preamble design and upstream burst-mode detection in 100-gb/s/$\lambda$ tdm coherent-pon,'' in \emph{Optical Fiber Communication Conference}.\hskip 1em plus 0.5em minus 0.4em\relax Optica Publishing Group, 2020, pp. W1E--1.

\bibitem{8}
J.~Zhang, Z.~Jia, H.~Zhang, M.~Xu, J.~Zhu, and L.~A. Campos, ``Rate-flexible single-wavelength tfdm 100g coherent pon based on digital subcarrier multiplexing technology,'' in \emph{2020 Optical Fiber Communications Conference and Exhibition (OFC)}.\hskip 1em plus 0.5em minus 0.4em\relax IEEE, 2020, pp. 1--3.

\bibitem{9}
D.~Welch, A.~Napoli, J.~B{\"a}ck, W.~Sande, J.~Pedro, F.~Masoud, C.~Fludger, T.~Duthel, H.~Sun, S.~J. Hand \emph{et~al.}, ``Point-to-multipoint optical networks using coherent digital subcarriers,'' \emph{Journal of Lightwave Technology}, vol.~39, no.~16, pp. 5232--5247, 2021.

\bibitem{xu2022intelligent}
M.~Xu, Z.~Jia, H.~Zhang, L.~A. Campos, and C.~Knittle, ``Intelligent burst receiving control in 100g coherent pon with 4$\times$ 25g tfdm upstream transmission,'' in \emph{Optical Fiber Communication Conference}.\hskip 1em plus 0.5em minus 0.4em\relax Optica Publishing Group, 2022, pp. Th3E--2.

\bibitem{konstadinidis2018multilayer}
C.~Konstadinidis, P.~Sarigiannidis, P.~Chatzimisios, P.~Raptis, and T.~D. Lagkas, ``A multilayer comparative study of xg-pon and 10g-epon standards,'' \emph{arXiv preprint arXiv:1804.08007}, 2018.

\end{thebibliography}
